\documentclass[10pt,superscriptaddress,twocolumn,amsmath,amssymb,aps,prb,reprint]{revtex4-2}

\usepackage{mathrsfs}
\usepackage{graphicx}
\usepackage{dcolumn}
\usepackage{bm}
\usepackage{color}

\usepackage{multirow}
\usepackage[colorlinks,bookmarks=false,citecolor=blue,linkcolor=red,urlcolor=blue]{hyperref}

\newcommand{\be}{\begin{equation}}
\newcommand{\ee}{\end{equation}}
\newcommand{\bea}{\begin{eqnarray}}
\newcommand{\eea}{\end{eqnarray}}

\definecolor{purple}{RGB}{128,0,128}

\begin{document}
\title{Time-Reversal Symmetry Breaking Superconducting State and Collective Modes in Kagome Superconductors}
\author{Xinloong Han}
\email{hanxinloong@gmail.com}
\affiliation{Department of Physics, Capital Normal University, Beijing 100048, China}
\affiliation{Kavli Institute for Theoretical Sciences, University of Chinese Academy of Sciences, Beijing 100190, China}

\author{Jun Zhan}
\affiliation{Beijing National Laboratory for Condensed Matter Physics and Institute of Physics, Chinese Academy of Sciences, Beijing 100190, China}
\affiliation{School of Physical Sciences, University of Chinese Academy of Sciences, Beijing 100190, China}
\author{Jiangping Hu}\email{jphu@iphy.ac.cn}
\affiliation{Beijing National Laboratory for Condensed Matter Physics and Institute of Physics, Chinese Academy of Sciences, Beijing 100190, China}
\affiliation{Kavli Institute for Theoretical Sciences, University of Chinese Academy of Sciences, Beijing 100190, China}
\author{Fu-chun Zhang}\email{fuchun@ucas.ac.cn}
\affiliation{Kavli Institute for Theoretical Sciences, University of Chinese Academy of Sciences, Beijing 100190, China}
 \author{Xianxin Wu}\email{xxwu@itp.ac.cn}
 \affiliation{Institute of Theoretical Physics,
Chinese Academy of Sciences, Beijing 100190, China}

\date{\today}
\begin{abstract}
We comprehensively study the unconventional pairing and collective modes in the multiband kagome superconductors AV$_3$Sb$_5$ (A=$\mathrm{K},\mathrm{Cs},\mathrm{Rb}$). By solving gap equations at zero temperature, we identify a transition from normal $s++/s\pm$-wave pairing to time-reversal symmetry (TRS) breaking pairing with a variation of inter-pocket interactions or density of states. This TRS breaking pairing originates from the superconducting phase frustration of different Fermi pockets and can account for experimental TRS breaking signal in kagome superconductors. Moreover, we investigate collective modes, including the Higgs, Leggett, and Bogoloubov-Anderson-Goldstone modes, arising from fluctuations of the amplitude, relative phase, and overall phase of the superconducting order parameters, respectively. Remarkably, due to the presence of multibands, one branch of the Leggett modes becomes nearly massless near the TRS breaking transition, providing a compelling smoking-gun signature of TRS-breaking superconductivity, in clear contrast to TRS-breaking charge orders. Our results elucidate the rich superconducting physics and its associated collective modes in kagome metals, and suggest feasible experimental detection of TRS breaking pairing.

\end{abstract}
 \maketitle

\section{Introduction}
The recently discovered kagome superconductors $A$V$_3$Sb$_5$ ($A$=K, Rb, Cs)\cite{oritz2019,oritz2020,oritz2o21PRM,oritz2021PRX} have attracted significant attention as a novel platform for studying exotic intertwined electronic orders because of its unique lattice structure and fascinating electronic features. Specifically, the interaction among sublattice interference effects associated with van Hove singularities~\cite{kiesel2012sublattice,wu2021PRL}, band topology and electronic correlations promotes intriguing rich phenomena. For example, theoretical calculations predict the existence of TRS breaking loop current state (imaginary charge density wave)\cite{feng2021chiral,feng2021low,jiang2023kagome,gu2022gapless}.  The prediction has gained much experimental supports\cite{jiang2021unconventional,guo2022switchable,mielke2022time}. Electronic nematicity\cite{kang2022twofold,nie2022charge,xiang2021twofold,li2023unidirectional,li2022rotation}, pair-density-wave\cite{chen2021roton,deng2024chiral}, double dome superconductivity (SC) under pressure\cite{Chen2021PRLdouble,chen2021highly,zhang2021pressure}, potential chiral superconductivity\cite{le2024superconducting,deng2024chiral,deng2024evidence} have also been observed. One of the main challenges in understanding the underlying pairing nature and competition between  charge orders and superconductivity in $A$V$_3$Sb$_5$ compounds, is to determine the  pairing symmetry and mechanism. On the theoretical side, calculations have suggested a variety of superconducting states, such as the chiral nodeless $d+i d$-wave singlet pairing\cite{yu2012chiral,qianghuawang2013,kiesel2012sublattice,lin2021complex,wu2021PRL,tazai2025nematic}, nodal $f$-wave triplet pairing\cite{Ronny2013,wu2021PRL} and $s_{\pm}$-wave pairing\cite{PhysRevB.108.L100510}. Meanwhile on the experimental side, many evidence have gradually aligned to support a nodeless\cite{duan2021nodeless,mu2021s,zhong2023nodeless}, nearly isotropic\cite{roppongi2023bulk} and singlet gap functions\cite{mu2021s}. Moreover, superconducting diode effect\cite{le2024superconducting}, scanning tunneling microscopy (STM) and muon spectral relaxation ($\mu$SR) measurements\cite{guguchia2023tunable,deng2024evidence} provide evidence of TRS breaking within the superconducting state when CDW order is suppressed. 
In pristine materials, $\mu$SR\cite{gupta2022microscopic}, point-contact spectroscopy\cite{yin2021strain}, angle-resolved photoemission spectroscopy (ARPES) and transport  measurements have identified two gap features, with one being anisotropic~\cite{2024arXiv240418472M,hu2024evidence,hossain2025unconventional}.

Density-functional theory (DFT) calculations and ARPES measurements on AV$_3$Sb$_5$ compounds reveal that the Fermi surfaces relevant to superconductivity originate not only from the V $d$ orbitals but also exhibit substantial contributions from the Sb $p$ orbitals~\cite{uykur2022_sb,Oey2022_sb,ritz2023superconductivity}, highlighting the important role of Sb in shaping the electronic structure. In addition, recent experimental evidence~\cite{zhong2023nodeless} suggests Cooper pairing between two hole pockets located at the $K$ and $K'$ valleys, further emphasizing the significance of valley-specific electronic states in the superconducting properties of AV$_3$Sb$_5$. Understanding the existence and contribution of the electron pocket from Sb atoms, as well as the pairing on the $d$-orbital dominated pockets, is essential for comprehending the overall behavior and properties of the superconducting state in kagome metals $A$V$_3$Sb$_5$. \begin{figure}[ht]
	\includegraphics[width=0.48\textwidth]{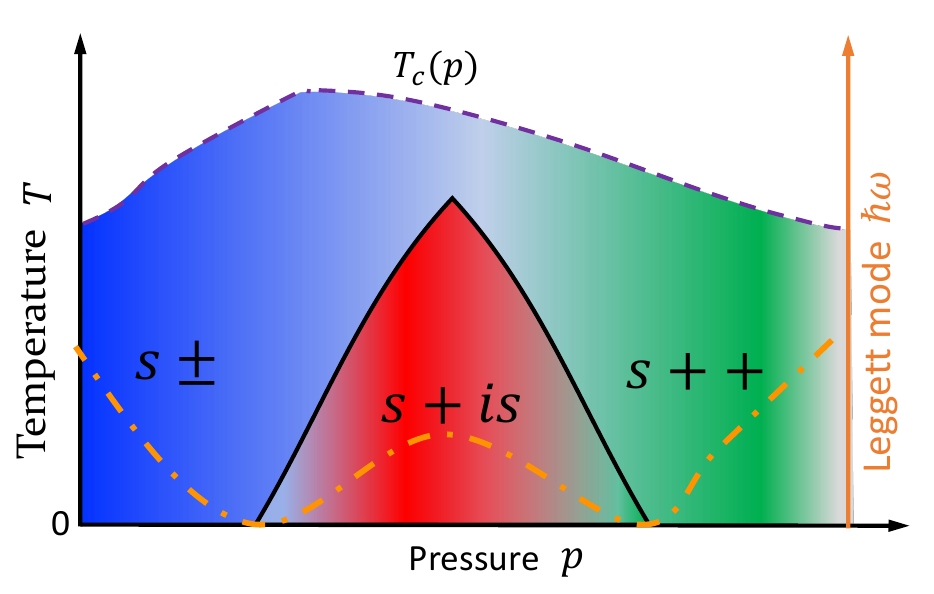}
	\caption{Schematic superconducting phase diagram with pressure $p$ and temperature $T$ from the multiband scenario in the kagome superconductors AV$_3$Sb$_5$. With increasing pressure (doping), the system undergoes a transition from a sign changing $s\pm$ to a trivial $s++$ state without sign change between two pockets below superconducting transition temperature $T_c(p)$. At the low temperature limit, there exists a time-reversal symmetry breaking superconducting state $s+is$ between the $s++$ and $s\pm$ state. In the vicinity of each superconducting transition, the spectrum of one branch of the Leggett mode exhibits a distinct minimum structure, which provides a characteristic signature for detecting TRS pairing in Kagome metals.
	}
	\label{fig:Fig1}
\end{figure}These findings strongly point to the multiband nature of superconductivity in this class of materials, which in turn has profound implications for the possible collective modes\cite{hu2024evidence,zhang2024higgs}, associated with the fluctuations of phase and amplitude of superconducting order parameter\cite{Anderson1958,Bogolyubov_1959,anderson1958random,varma1981gauge,varma1982,pekker2015amplitude,leggett1966number,xiaohu2012massless_prl,chubukov2012s+is_prb,Stanev2012,Benfatto2013Leggettiron_prb,Benfatto2016_prb}. These collective modes also provide a useful framework to analyze such multiband superconductors. The intriguing mode called the Leggett mode\cite{leggett1966number} can arise from fluctuations in the relative phase between superconducting phases in different bands, with its mass proportional to the interband coupling strength. Typically, the massive Leggett mode undergoes decay due to two-quasiparticles continuum. However, in certain scenarios, the Leggett mode can become massless in the TRS breaking superconducting phase\cite{xiaohu2012massless_prl,chubukov2012s+is_prb,Stanev2012,Benfatto2013Leggettiron_prb,Benfatto2016_prb}, leading to its stabilization.

\begin{figure*}[ht]
	\includegraphics[width=0.75\textwidth]{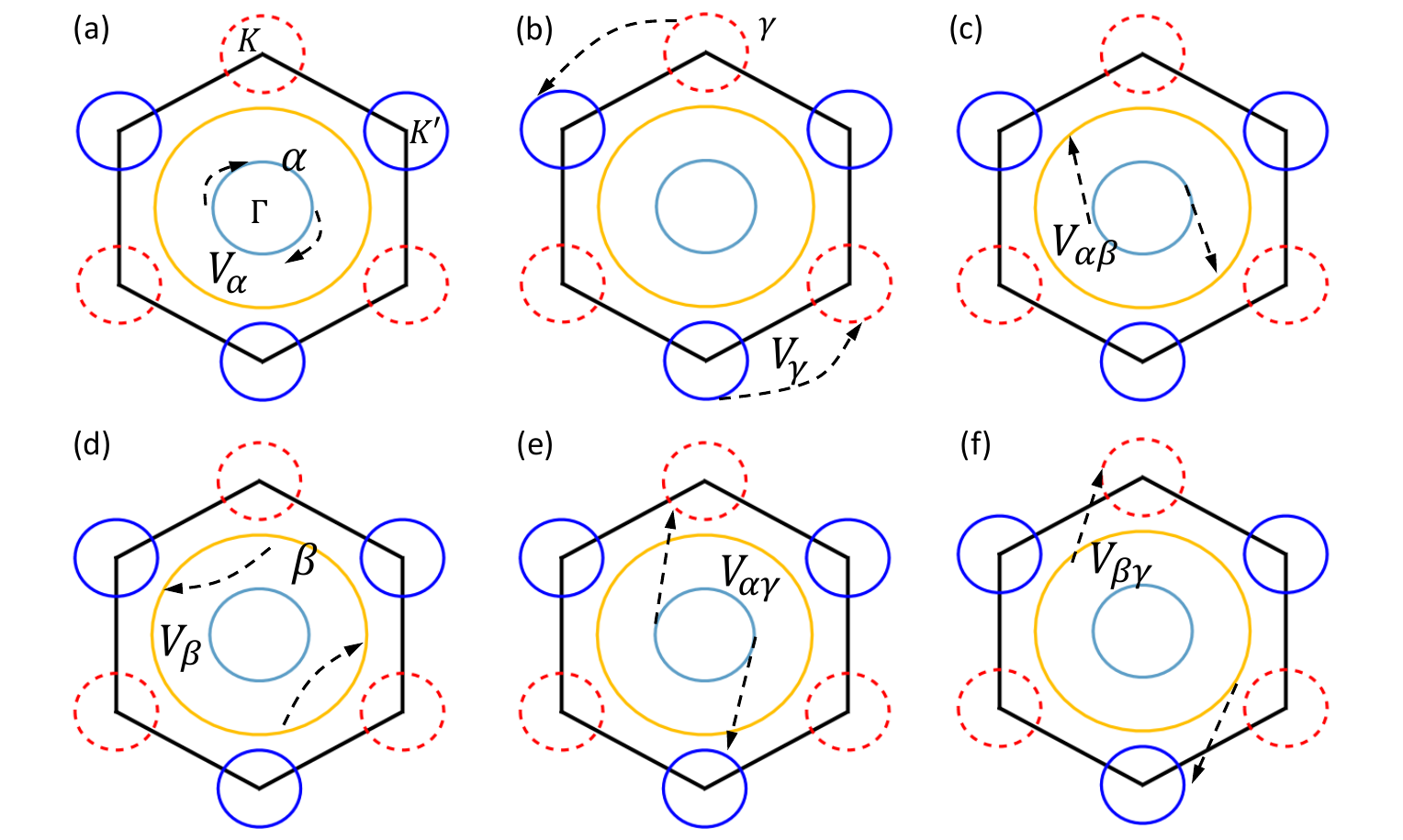}
	\caption{Scattering processes of the four Fermi pockets as denoted by $\alpha$, $\beta$ and $\gamma$, respectively. Intra-pockets interactions $V_{\alpha}$ (a) and $V_{\beta}$ (d) and inter-pocket interaction $V_{\gamma}$ (b) between electrons at different valleys. (c) The pair-hopping process from $\alpha$ pocket contributed by the $p$ orbital of Sb atoms to $\beta$ pocket contributed by $d$ orbitals of V atoms. We set $V_{\alpha\beta}=-u_{\alpha\beta}/V$. (e) The pair-hopping process from $\alpha$ pocket to $\gamma_{K,K^{\prime}}$ pocket contributed by $d$ orbitals of V atoms. (f) The pair-hopping process from $\beta$ pocket to $\gamma_{K,K^{\prime}}$.
	}
	\label{fig:Fig2}
\end{figure*}

Motivated by above observations, in this paper we perform a comprehensive study on superconducting pairing symmetry and collective modes based on the scenario of  multiband nature of kaogme superconductors. Oue main results are illustrated in Fig. \ref{fig:Fig1}, in which there manifest a TRS breaking superconducting state and one branch of Leggett mode demonstrate minina structures in its spectra. The reminder of this paper is organized as follows. In the section \ref{sec:model}, we introduce our model and derive the gap equation. In the section \ref{sec:Emergence of TRSB}, we present the numerical results to reveal the existence of $s+is$ superconducting phase with TRS breaking. In section \ref{sec:Collectivemodes}, we further study the collective modes. Finally, in the section \ref{sec:conclusion} we give our further discussions and conclude this paper.

\begin{figure*}[ht]
	\includegraphics[width=0.88\textwidth]{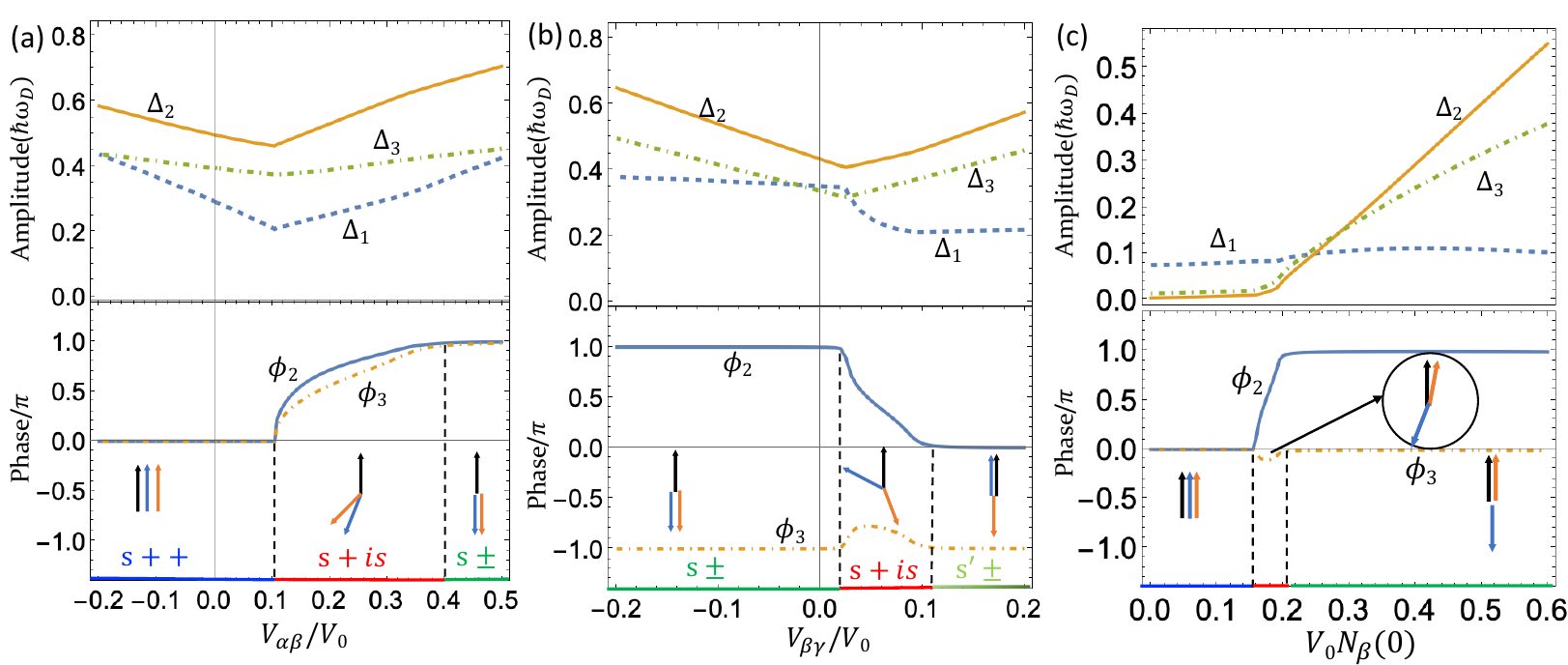}
	\caption{
	The evolution of amplitudes and phases of the $\Delta_{\alpha,\beta,\gamma}$ as changing inter-pocket scattering strength or DOS $V_0N_\beta(0)$ with initial intrapocket interaction as $V_{\alpha,\beta,\gamma}=-V_0$. (a) Superconducting state evolves from $s++$ to $s\pm$ as increasing $V_{\alpha\beta}$. In the middle region with repulsive interaction, there exists a TRS breaking $s+is$ state. The initial parameters are chosen as $V_{\alpha\gamma}= V_{\beta\gamma}=-0.1V_0$. (b) As increasing $V_{\beta\gamma}$, there exists a phase transition from $s\pm$ to TRS breaking state. When the repulsive interaction between $\beta$ and $\alpha$ pocket is strong enough, the phase goes into another type of $s\pm$ state, which is denoted by $s^{\prime}\pm$ marked by light green line. The initial parameters are chosen as $V_{\alpha\beta}=V_{\alpha\gamma}=0.1V_0$. In (a) and (b), we set $V_0N_{\alpha,\gamma}(0)=0.4$ and $V_0N_{\beta}(0)=0.6$. (c) By tuning DOS of $\beta$ pocket $V_0N_{\beta}(0)$, a TRS breaking state separate $s++$ and $s\pm$ state; we chose $V_0N_{\alpha,\gamma}(0)=0.3$, $V_{\alpha\beta,\alpha\gamma}=0.1V_0$ and $V_{\beta\gamma}=0.2V_0$. In these figures, blue, red and green line represents the region of $s++$, $s+is$ and $s\pm$ superconducting state, respectively. Here we only illustrate one typical solution of $\phi_{2,3}$, and other solutions $\tilde{\phi}_{2,3}$ can be obtained from $\cos(\tilde{\phi}_{2,3})=\cos(\phi_{2,3})$ and $\cos(\tilde{\phi}_2-\tilde{\phi}_3)=\cos(\phi_2-\phi_3)$.
	}
	\label{fig:Fig3}
\end{figure*}

\section{The model}\label{sec:model}

In this section, we construct phenomenological model to study the pairing states for kagome superconductors based on fermiology in experiments. 
Recent ARPES measurements on slightly doped CsV$_3$Sb$_5$\cite{zhong2023nodeless} have identifies superconducting gaps on three Fermi surface sheets: (1) A large $\Gamma$-centered hole pocket $\beta$ originating from V $d$-orbitals; (2) an additional electron pocket, from the contribution of V $d$-orbitals, situated at $K$ or $K^{\prime}$ valley; (3) a $\Gamma$-centered electron pocket emerging from Sb $p$-orbital.
We neglect the detailed shape of Fermi surfaces and consider four round pockets in our model, as shown in Fig. \ref{fig:Fig2}. The two pockets around the $\Gamma$ point are labelled $\alpha$ and $\beta$, and two pockets around the $K$ point are labelled $\gamma$,  where we use $\gamma_{K}$, and $\gamma_{K^{\prime}}$ to distinguish $\gamma$ pockets at $K$ and $K^\prime$ . Since the zero-momentum pairing mechanism is considered, the effective model can be cast into,
\bea
&&S=\int d\tau d^2{\bf r} \Big\{\sum_{i,\sigma}\bar{\psi}_{i\sigma}(\tau,{\bf r})(\partial_{\tau}+\epsilon_{i}(-i\nabla_{\bf r})-\mu_{j})\psi_{i\sigma}(\tau,{\bf r})\nonumber \\
&&-\sum_{jl}\hat{g}_{jl} \bar{\chi}_{j}(\tau,{\bf r}) \chi_{l}(\tau,{\bf r})\Big\},
\eea
where $\bar{\psi}_{i\sigma}$ fermionic field means creating a electron (hole) with spin $\sigma=\uparrow,\downarrow$ at the $i$ ($i=\alpha,\beta,\gamma_{K},\gamma_{K^{\prime}}$) pocket. $\epsilon_i(-i\nabla)$ is the dispersion of electrons (holes) near $i$ pocket. $\mu_j$ is the chemical potential; In this paper, we set $\mu_{\alpha,\beta}=\mu$ for electron pockets and $\mu_{\gamma_K,\gamma_{K^{\prime}}}=-\mu$ for hole pockets.  Here we take $\chi_j$ as pairing operator, and they are given as $\chi_{\alpha}=\psi_{\alpha\downarrow} \psi_{\alpha\uparrow}$, $\chi_{\beta}=\psi_{\beta\downarrow} \psi_{\beta \uparrow}$, and $\chi_{\gamma_K}=\psi_{\gamma_K\downarrow} \psi_{\gamma_{K^{\prime}}\uparrow}$, and $\chi_{\gamma_{K^{\prime}}}=\psi_{\gamma_{K^{\prime}}\downarrow} \psi_{\gamma_{K}\uparrow}$. In general the interactions $\hat{g}$ can be taken as,
\bea
\hat{g}=-\left(\begin{array}{cccc}
V_\alpha & V_{\alpha\beta}  & V_{\alpha\gamma} &V_{\alpha\gamma} \\
V_{\alpha\beta} & V_{\beta} &V_{\beta\gamma} &V_{\beta\gamma} \\
V_{\alpha\gamma} & V_{\beta\gamma} & V_{\gamma} & 0 \\
V_{\alpha\gamma} &V_{\beta\gamma} & 0 & V_{\gamma}
\end{array}
\right).
\eea
Here $V_{\alpha,\beta,\gamma}$ represents intrapocket interaction, $V_{\alpha\beta}$, $V_{\alpha\beta}$ and $V_{\beta\gamma}$ is the pair hoping term between different pockets. The interactions $\hat{g}$ satisfies $\mathrm{Det}(\hat{g})>0$. Hereafter it is well-defined to perform Hubbard-Stratonovich (HS) transformation to decouple interaction term by introducing pairing fields as $\Delta_{i}$,
\bea
&&S[\Delta,c]=\int d\tau d^2{\bf r} \Big\{\sum_{i,\sigma}\bar{\psi}_{i\sigma}(\tau,{\bf r})(\partial_{\tau}+\epsilon_{i}(-i\nabla_{\bf r})-\mu_j)\nonumber \\
&&\times \psi_{i\sigma}(\tau,{\bf r})-\sum_{j}\big(\Delta_{j} \chi^{\dagger}_{j}+\Delta^{\dagger}_j \chi_j-\sum_{l}\hat{g}^{-1}_{jl}\Delta_{j}^{\dagger} \Delta_l\big)\Big\}.
\eea

After integrating out the fermionic fields, we can obtain
\bea
S[\Delta]=-\sum_{i=\alpha,\beta,\gamma}tr \ln G^{-1}_{i}+\int d\tau d^2{\bf r}\sum_{jl}\hat{g}^{-1}_{jl}\Delta_{j}^{\dagger} \Delta_l,
\eea
with Green's function $G_{i=\alpha,\beta}(\tau,{\bf r})$ defined in the Nambu space $(\psi_{i\uparrow},\bar{\psi}_{i\downarrow})$ for $\alpha$ or $\beta$ pocket,
\bea
G_i^{-1}=\left( \begin{array}{cc}
-\partial_{\tau}-\varepsilon_i(-i\nabla_{\bf r}) & \Delta_{i}\\
\Delta_i^{\dagger} & -\partial_{\tau}+\varepsilon_i(i\nabla_{\bf r})
\end{array}\right),
\eea
where $\varepsilon_i(-i\nabla_{\bf r})=\epsilon_i(-i\nabla_{\bf r})-\mu$. For the fermionic fields in $\gamma_{K,K^{\prime}}$ pocket, the Green's function $G_{\gamma}$ is defined in the valley and Nambu space $(\psi_{\gamma_K\uparrow},\bar{\psi}_{\gamma_{K^{\prime}}\downarrow},\psi_{\gamma_{K^{\prime}}\uparrow},\bar{\psi}_{\gamma_{K}\downarrow})$,
\bea
G^{-1}_{\gamma}=\left(\begin{array}{cc}
G^{-1}_{\gamma_{K^\prime}} (\tau,{\bf r}) &0 \\
0 &G^{-1}_{\gamma_K}(\tau,{\bf r})
\end{array}\right)
\eea
with 
\bea
G^{-1}_{\gamma_{K,K^{\prime}}}=\left(\begin{array}{cc}
-\partial_{\tau}+\varepsilon_{\gamma_{K,K^\prime}}(-i\nabla_{\bf r}) & \Delta_{\gamma_{K,K^{\prime}}} \\
\Delta_{\gamma_{K,K^{\prime}}}^{\dagger} & -\partial_{\tau}-\varepsilon_{\gamma_{K^{\prime},K}}(-i\nabla_{\bf r})
\end{array}\right).\nonumber \\
\eea

Based on the above effective action for pairing fields $\Delta_i$ and $\Delta^{\dagger}_i$, we can obtain the following gap equations for superconductivity from saddle-point equations $\partial S[\Delta]/\partial \Delta_j^{\dagger}=0$ at zero temperature as,
\bea\label{eq:gapeq}
\Delta_j= \sum_{jl} \hat{g}_{jl} N_l(0)\Delta_l \mathrm{arsinh}(\frac{\hbar \omega_D}{|\Delta_{l}|}),
\eea
where $\hbar\omega_D$ is taken as an energy cutoff for Debye frequency or spin fluctuation energy scale. $N_j(0)$ represents the density of states (DOS) near the FS for $j$ ($j=\alpha,\beta,\gamma_K,\gamma_{K^{\prime}}$) pocket.  Furthermore, the free-energy density $\mathcal{F}[\Delta]$ at the saddle-point with zero temperature is given by,
\bea\label{eq:freeeng}
&&\mathcal{F}[\Delta]=\sum_{jl}\hat{g}^{-1}_{jl}\Delta_j^{\dagger}\Delta_l-\sum_jN_j(0)|\Delta_j|^2\nonumber \\
&&\times\big[\frac{1}{2}+\mathrm{arsinh}(\frac{\hbar \omega_D}{|\Delta_{j}|})\big].
\eea Together with gap equations, we can obtain the ground state of the system. Note the system has $\hat{C}_6$ symmetry, thus we have $\Delta_{\gamma_K}=\Delta_{\gamma_{K^{\prime}}}$. We take $\Delta_{\alpha}$ to be $\Delta_1e^{i\phi_1}$ as a fixed gauge of the system as $\phi_1=0$. Meanwhile, we rewrite $\Delta_{\beta}=\Delta_2 e^{i\phi_2}$ and $\Delta_{\gamma_K,\gamma_{K^\prime}}=\Delta_3 e^{i\phi_3}$, with $\Delta_{2,3}>0$.


\section{Emergence of TRS breaking $s+is$ state}\label{sec:Emergence of TRSB}
In this section, we systematically study the pairing symmetry for different inter-packet scattering processes originating from electron-phonon coupling or spin/charge fluctuations. The newly angle-resovled photoemission spectroscopy measurements\cite{luo2022electronic,zhong2023testing} revealed kinks in both V $d$-orbital and Sb $p$-orbital bands, suggesting an electron-phonon coupling strength. Hence, it is reasonable to assume intrapocket pairing interactions are attractive. For simplicity, we set $V_{\alpha,\beta,\gamma}=-V_0$ in the rest of the paper, where $V_0$ is the unit interaction strength with $V_0>0$. Due to electron phonon coupling and intrinsic spin/charge fluctuations in kagome superconductors, we take interpocket pairing scattering as tuning parameters varying from attractive from repulsive in the following calculations.

By numerically solving gap equations Eq. \ref{eq:gapeq} and minimizing Eq. \ref{eq:freeeng}, we firstly demonstrate paring gap structures as a function of inter-pocket interaction $V_{\alpha\beta}$ as shown in Fig. \ref{fig:Fig3}(a), with chosen the electron-phonon dominated interactions as  $V_{\beta\gamma}=V_{\alpha\gamma}=-0.1V_0$, and larger DOS for $\beta$ pocket as $V_0N_{\beta}(0)=0.6$ and smaller ones as $V_0 N_{\alpha,\gamma}(0)=0.4$. The attractive interaction $V_{\alpha\gamma}$ and $V_{\beta\gamma}$ favor the same phase for superconducting gaps $\Delta_{\alpha,\beta,\gamma}$, leading to $\phi_2=\phi_3=\phi_1$. A repulsive interaction $V_{\alpha\beta}$ favors $\phi_2=\phi_1\pm \pi$, inducing phase frustration in $\phi_{1,2,3}$. When the repulsive interaction $V_{\alpha\beta}$ is significant, the strong phase frustration will generate a time-reversal symmetry breaking superconducting state with $\phi_{2,3}\neq 0,\pi$, resulting $\Delta_{\beta,\gamma}\neq \Delta_{\beta,\gamma}^*$. As shown in the Fig. \ref{fig:Fig3}(a), the transition from $s++$ to a TRS breaking $s+is$ state occurs around $V_{\alpha\beta}=0.1V_0$. With further increasing $V_{\alpha\beta}$, $\phi_{2,3}$ will increase gradually. When the repulsive interaction $V_{\alpha\beta}$ is much stronger than $V_{\alpha\gamma,\beta\gamma}$, i.e. $V_{\alpha\beta}>0.4V_0$, the $s\pm$ pairing symmetry with $\phi_{2,3}=\pm \pi$ is favored, as shown in the Fig. \ref{fig:Fig3}(a).

As $\beta$ and $\gamma$ pockets are mainly attributed to V $d$ orbitals, the corresponding pair scattering $V_{\beta\gamma}$ can be tuned by doping or pressure. Fig. \ref{fig:Fig3}(b) displays the evolution of gap structures a function of the $V_{\beta\gamma}$, calculated with repulsive interpocket interactions fixed at $V_{\alpha\beta}=V_{\alpha\gamma}=0.1 V_0$. The fact that repulsive $V_{\alpha\beta,\alpha\gamma}$ favors $\phi_{2,3}=\phi_1\pm\pi$ and attractive interaction $V_{\beta\gamma}$ will lock the $\phi_{2}$ and $\phi_3$ to be equal, leads to the $s\pm$ superconducting state, with sign change of the gap between $\alpha$ and $\beta$ or $\gamma$ pocket. However, a repulsive $V_{\beta\gamma}$ will induce phase frustration in these superconducting gaps. If  $V_{\beta\gamma}$ is dominant, another type of $s\pm$ with a sign change of the superconducting gaps between $\gamma$ and $\alpha$ or $\beta$ pocket. The pairing is denoted as $s^{\prime}\pm$. For moderate $V_{\beta\gamma}$, a TRS breaking $s+is$ superconducting state emerge, separating these two kinds of $s\pm$ states, as illustrated in Fig. \ref{fig:Fig3}(b).

The DOS can also play a critical role in shaping the characteristics of the superconducting state. In the kagome metal AV$_3$Sb$_5$, it is noteworthy that DOS originating from $\beta$ pocket, primarily contributed by V $d$-orbitals, surpass that of other pockets. This elevated DOS of $\beta$ pocket finds its origin in the presence of multiple VHSs\cite{kang2022twofold,hu2022rich}. Under the influence of doping or external pressure, the DOS associated with $\beta$ pocket can be manipulated by tuning the proximity between the Fermi level and the VHSs. To elucidate this influence, our investigation delves into the pairing symmetry, specifically by modulating the DOS of the $\beta$ pocket.
As depicted in Fig. \ref{fig:Fig3}(c), our calculations reveal a gradual evolution of the superconducting state with varying $N_{\beta}(0)$. Initially displaying characteristics akin to an $s++$ pairing symmetry, the system transitions through intermediate stages, eventually culminating in a TRS breaking state. Intriguingly, with further increments in the parameter $V_0N_\beta(0)$, the superconducting state ultimately evolves to the $s\pm$ pairing symmetry.

Therefore, tuning electronic interactions and DOS can result a TRS breaking pairing based on the multi-pocket model. The microscopic origin lies in the phase frustration of pairings on different bands, analogous to iron based superconductors~\cite{Chubukov2013}. The proposed pairing state may account for the observed nodeless pairing with TRS breaking signal in kagome superconductors when CDW is eliminated by with external pressure or doping~\cite{guguchia2023tunable,zhong2023nodeless,deng2024evidence}.

\section{Collective modes: BAG, Leggett and Higgs modes}\label{sec:Collectivemodes}

In this section, we explore collective modes of kagome superconductors. To evaluate the fate of collective modes during the transition of TRS breaking state, we investigate fluctuations of phase and amplitude of superconducting order parameters. By investigating these modes, it is convenient to  write order parameters as\cite{Melo1997,Mohit2008,han2016} $\Delta_{i}(x)=\Delta_{i}^0(1+\delta h_{i}) e^{i\theta_i}$ in which $\delta h_i$ describes amplitude fluctuation. In addition, we introduce the $U(1)$ gauge transformation\cite{loktev2001phase,Mohit2008,xiaohu2012massless_prl} as $\psi_j\rightarrow \psi_je^{i\theta_j/2}$, $\bar{\psi}_j \rightarrow \bar{\psi}_je^{-i\theta_j/2}$, $\Delta_j\rightarrow \Delta_j e^{i\theta_j}$ for $j=\alpha,\beta$ pocket and $\Delta_j e^{i(\theta_{\gamma_K}+\theta_{\gamma_{K^{\prime}}})/2}, j=\gamma_K, \gamma_{K^{\prime}}$ for $j=\gamma_{K},\gamma_{K^{\prime}}$.
The $U(1)$ gauge transformation and $\delta h_i$ will give corrections to the effective action beyond the saddle-point solutions. This point is reflected in the Green's function by introducing a self-energy $\Sigma_{j,\theta}$ and $\Sigma_{j,h}$ as $\mathcal{G}_j^{-1}=G_{j}^{-1}-\Sigma_{j,\theta}-\Sigma_{j,h}$. For the two electron pockets $j=\alpha,\beta$, $\Sigma_{j,\theta}$ can be written as
\bea
\Sigma_{j,\theta}=\Big[\frac{i}{2}\partial_{\tau}\theta_j+\frac{(\nabla \theta_j)^2}{8m}\Big]\hat{\sigma}_3-\frac{i}{2m_j}\Big[\frac{\nabla^2 \theta_j}{2}+ \nabla\theta_j\nabla \Big]\hat{\sigma}_0,\nonumber \\
\eea
where $\hat{\sigma}_0$ is the unit $2\times2$ matrix and $\hat{\sigma}_3$ is the Pauli matrix. For the two hole $\gamma$ pockets connected by inversion symmetry,  the self-energy is slightly different from  the $\alpha,\beta$ pockets. In the Nambu space $(c_{j\uparrow}^{\dagger}, c_{\bar{j}\downarrow})$ for hole pockets $j=\gamma_{K},\gamma_{K^{\prime}}$, the self-energy $\Sigma_{j,\theta}$ is given by
\bea
&\Sigma_{j,\theta}=\Big[\frac{i}{2}\partial_{\tau}\theta_{j}+\frac{(\nabla \theta_{j})^2}{8m}-\frac{i}{2m_j}\big(\frac{\nabla^2 \theta_{j}}{2}+ \nabla\theta_{j}\nabla \big)\Big]\hat{\sigma}_+\nonumber \\
&+\Big[-\frac{i}{2}\partial_{\tau}\theta_{\bar{j}}-\frac{(\nabla \theta_{\bar{j}})^2}{8m}-\frac{i}{2m_{j}}\big(\frac{\nabla^2 \theta_{\bar{j}}}{2}+ \nabla\theta_{\bar{j}}\nabla \big)\Big]\hat{\sigma}_-,
\eea
where $\bar{\gamma_K}=\gamma_{K^{\prime}}$ and $\bar{\gamma_{K^{\prime}}}=\gamma_{K}$. Here we define $\hat{\sigma}_{\pm}=(\hat{\sigma}_{0}\pm\hat{\sigma}_{3})/2$. And the self energy $\Sigma_{j,h}$ contributed by the fluctuations of amplitudes are 
\bea
\Sigma_{j,h}(x)=-\Delta_{j}^0 \delta h_j\left(\begin{array}{cc}
0 & e^{i\tilde{\theta}_j} \\
e^{-i\tilde{\theta}_j} & 0
\end{array}
\right),
\eea
with $\tilde{\theta}_{\alpha,\beta}=\theta_{\alpha,\beta}$ and $\tilde{\theta}_{\gamma_K,\gamma_{K^\prime}}=(\theta_{\gamma_K}+\theta_{\gamma_{K^\prime}})/2$. These phases of superconducting order parameters will oscillate around saddle points as $\tilde{\theta}_{\alpha,\beta}=\phi_{1,2}+\delta \phi_{1,2}$ and $\tilde{\theta}_{\gamma_K,\gamma_{K^\prime}}=\phi_3+\delta\phi_3$. 
\begin{figure*}
	\includegraphics[width=0.8\textwidth]{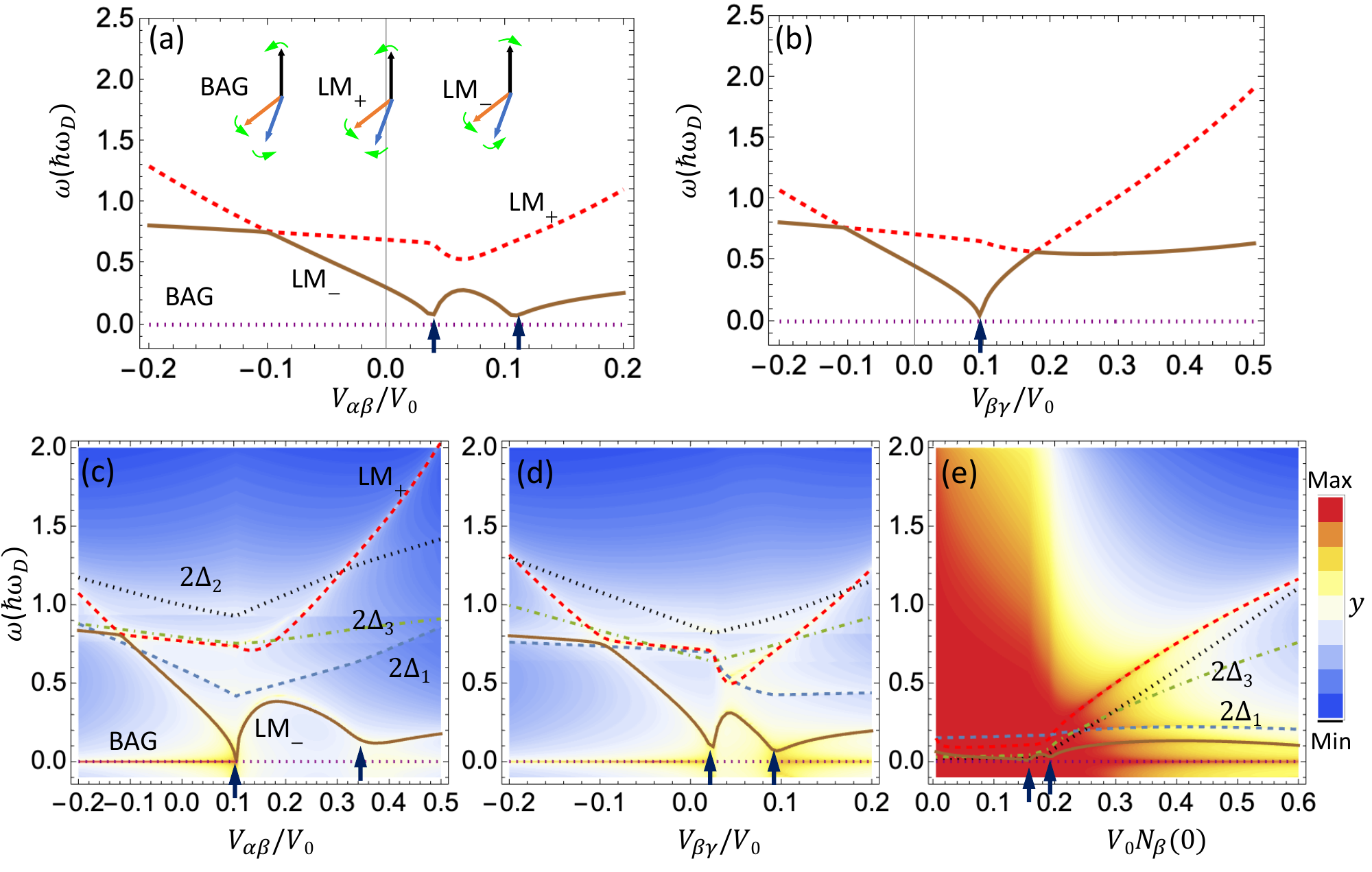}
	\caption{Collective modes as functions of interpocket interaction or DOS at the zero momentum ${\bf q}=0$. (a) and (b) illustrate the evolution of masses of BAG and two LMs with increasing interpocket interaction $V_{\alpha\beta}$, and $V_{\beta\gamma}$, respectively. Green arrows represent relative oscillations of the corresponding phase. Here we set $V_0N_{\alpha,\beta,\gamma}=0.5$. (c,d,e) indicates the collective modes resolved from solving $y=-\ln(|\mathrm{Det}{\bf Q}(\omega,0)|)$ as varing $V_{\alpha\beta}$, $V_{\beta\gamma}$ or $V_0N_{\beta}(0)$. Here we plot two-quasiparticel continuum setted by $2\Delta_{1,2,3}$. In (a) and (c), we set parameters $V_{\alpha\gamma}=V_{\beta\gamma}=-0.1V_0$; in (b) and (d) we set $V_{\alpha\beta}=V_{\alpha\gamma}=0.1V_0$. In (c) and (d), we also have $V_0N_{\alpha,\gamma}(0)=0.4$ and $V_0N_{\beta}(0)=0.6$. In (e), we chose $V_0N_{\alpha,\gamma}(0)=0.3$, $V_{\alpha\beta,\alpha\gamma}=0.1V_0$ and $V_{\beta\gamma}=0.2V_0$. The solid black arrows indicate the TRS breaking transition points where mass of LM $\Omega_{\mathrm{LM}-}$ becomes vanishing small.
	} 
	\label{fig:Fig4}
\end{figure*}
After integrating out Fermionic fields and expanding to the second order of $\delta h$ and $\delta \phi$, we conclude the following effective theory for collective modes as,
\bea
&&S_{cm}=\frac{1}{8}\sum_{i\Omega_l,{\bf q}}\bar{\Phi}(-i\Omega_l,-{\bf q}) \left(\begin{array}{cc}
{\bf O}^{hh} & {\bf O}^{h\phi} \\
({\bf O}^{h\phi})^{\dagger} & {\bf O}^{\phi\phi}
\end{array}
\right)
\Phi(i\Omega_l,{\bf q})\nonumber \\
&&=\frac{1}{8}\sum_{i\Omega_l,{\bf q}}\bar{\Phi}(-i\Omega_l,-{\bf q})\hat{{\bf O}}\Phi(i\Omega_l,{\bf q}),
\eea
with $\bar{\Phi}({i\Omega_l,q})=\left(\delta h_i(i\omega_l,q),\delta \phi_i(i\omega_l,q)\right)$ and ${\bf O}_{hh,h\phi,\phi\phi}$ is the $3\times 3$ matrix which is given in the long wavelength limit ${\bf q}\rightarrow {\bf 0}$ as followings,
\begin{widetext}
\bea
&&{\bf O}^{hh}=\left(\begin{array}{ccc}
8g^{-1}_{11}|\Delta_1^0|^2+4B_1(i\Omega_l,{\bf q}) & 8 g_{12}^{-1}|\Delta_1^0| |\Delta_2^0| \cos(-\phi_2) & 16 g_{13}^{-1}|\Delta_1^0| |\Delta_3^0| \cos(-\phi_3) \\
8 g_{12}^{-1}|\Delta_1^0| |\Delta_2^0| \cos(-\phi_2) & 8g^{-1}_{22}|\Delta_2^0|^2+4B_2(i\Omega_l,{\bf q}) & 16 g_{23}^{-1}|\Delta_2^0| |\Delta_3^0| \cos(\phi_2-\phi_3) \\
16 g_{13}^{-1}|\Delta_1^0| |\Delta_3^0| \cos(-\phi_3) &16 g_{23}^{-1}|\Delta_2^0| |\Delta_3^0| \cos(\phi_2-\phi_3)  & 16g^{-1}_{33}|\Delta_3^0|^2+8B_3(i\Omega_l,{\bf q})
\end{array}
\right), \\
&&{\bf O}^{h\phi}=\left(\begin{array}{ccc}
 8{\bf O}^{h\phi}_{11}& -8g^{-1}_{12} |\Delta_1^0| |\Delta_2^0|\sin(-\phi_2) & -16g^{-1}_{13} |\Delta_1^0| |\Delta_3^0|\sin(-\phi_3) \\
8g^{-1}_{12} |\Delta_1^0| |\Delta_2^0|\sin(-\phi_2) & 8{\bf O}^{h\phi}_{22} &-16 g^{-1}_{23}|\Delta_2^0| |\Delta_3^0| \sin(\phi_2-\phi_3) \\
16 g^{-1}_{13} |\Delta_1^0| |\Delta_3^0|\sin(-\phi_3) & 16 g^{-1}_{23} |\Delta_1^0| |\Delta_3^0|\sin(\phi_2-\phi_3) & 8{\bf O}^{h\phi}_{33}
\end{array}
\right),\\
&&{\bf O}^{\phi\phi}=\left(\begin{array}{ccc}
N_{\alpha}(0)(-2\Omega_l^2+v_{\alpha}^2|q|^2)-C_{1}&8 g_{12}^{-1}|\Delta_1^0| |\Delta_2^0| \cos(-\phi_2) & 16 g_{13}^{-1}|\Delta_1^0| |\Delta_3^0| \cos(-\phi_3) \\
8 g_{12}^{-1}|\Delta_1^0| |\Delta_2^0| \cos(-\phi_2) & N_{\beta}(0)(-2\Omega_l^2+v_{\beta}^2|q|^2)-C_{2} &16 g_{23}^{-1}|\Delta_2^0| |\Delta_3^0| \cos(\phi_2-\phi_3) \\
16 g_{13}^{-1}|\Delta_1^0| |\Delta_3^0| \cos(-\phi_3) & 16 g_{23}^{-1}|\Delta_2^0| |\Delta_3^0| \cos(\phi_2-\phi_3) & N_{\gamma}(0)(-4\Omega_l^2+2v_{\gamma}^2|q|^2)-C_{3}
\end{array}
\right),
\eea
\end{widetext}
where ${\bf O}^{h\phi}_{11}=2g^{-1}_{12}|\Delta_1^0| |\Delta_2^0| \sin(-\phi_2)+4g^{-1}_{13}|\Delta_1^0| |\Delta_3^0| \sin(-\phi_3)$, ${\bf O}^{h\phi}_{22}=-2g^{-1}_{12}|\Delta_1^0| |\Delta_2^0| \sin(-\phi_2)$ and ${\bf O}^{h\phi}_{33}=-4g^{-1}_{13}|\Delta_1^0| |\Delta_3^0| \sin(-\phi_3)$. We also have $C_1={\bf O}^{h\phi}_{12}+{\bf O}^{h\phi}_{13}$, $C_2={\bf O}^{h\phi}_{12}+{\bf O}^{h\phi}_{23}$ and $C_3={\bf O}^{h\phi}_{13}+{\bf O}^{h\phi}_{23}$. Here $B_{i}$ is defined as
\bea
B_i(i\Omega_l,{\bf q})=\frac{2|\Delta_i^0|^2}{V}\sum_{\bf k}\frac{E_i+E_i^{\prime}}{2E_iE_i^{\prime}}\frac{E E_i^{\prime}+\varepsilon_i\varepsilon_i^{\prime}-|\Delta_i|^2}{(i\Omega_l)^2-(E_i+E_i^{\prime})^2},
\eea
where $\varepsilon_i^{\prime}=\varepsilon_i({\bf k}+{\bf q})$, $\varepsilon_i=\varepsilon_{i}({\bf k})$, $E_{i}^{\prime}=\sqrt{|\varepsilon_i^{\prime}|^2+|\Delta_i|^2}$ and $E_{i}=\sqrt{|\varepsilon_i|^2+|\Delta_i|^2}$. The size of the system is denoted as $V$. Let us now discuss the general structure of $\hat{\bf O}$. First, the matrix ${\bf O}^{h\phi}$, which describes the interaction between phase and amplitude degrees of freedom, becomes vanishing in the non-TRS breaking phase with $\phi_{2,3}=0,\pm \pi$. Second, the amplitude degrees of freedom in different bands couple with each other in both the non-TRS and TRS breaking states, since off-diagonal elements of ${\bf O}^{hh}$ are proportional to $\cos(\phi_{2,3})$ or $\cos(\phi_2-\phi_3)$. In the TRS breaking state, fluctuations of phase and amplitude couple and the corresponding collective modes can be identified from the condition $\mathrm{Det}\hat{{\bf O}}=0$.

To develop an intuitive understanding of the collective modes linked to phase fluctuations, we initially disregard their intermingling with Higgs modes. This approximation holds in the non-TRS breaking phase and remains valid at the boundaries separating the TRS and non-TRS breaking phases, where the sine functions in ${\bf O}^{h\phi}$ render such contributions negligible. By solving the equation $\mathrm{Det}{\bf O}^{\phi\phi}(\Omega,0)=0$ 
with setting $N_{\alpha,\beta,\gamma}(0)=N_F(0)$ and $v_{\alpha,\beta,\gamma}=v_F$, we derive three distinct modes. Among these, one manifests as a massless BAG mode, characterized by the dispersion relation,

\bea
\Omega_{\mathrm{BAG}}^2=\frac{v_F^2 q^2}{2}, 
\eea
with associated with overall phase fluctuation. The massless BAG mode transforms to a plasma mode as $\Omega^2_{\mathrm{PM}}=v_F^2q^2(1+2N_F(0)V_{col}(q))/2$ by the so-called Anderson-Higgs mechanism, through coupling with charge fluctuation. The coulomb interaction in two dimension is $V_{col}(q)=2\pi/(\epsilon_B |q|)$ where $\epsilon_B$ is the dielectric constant. The other two are Leggett modes with dispersion 
\bea
&&\Omega_{\mathrm{LM}+}^2=\frac{v_F^2 q^2}{2}+\Delta_{\mathrm{LM}+}^2, \\
&&\Omega_{\mathrm{LM}-}^2=\frac{v_F^2 q^2}{2}+\Delta_{\mathrm{LM}-}^2,\label{eq:LM-}
\eea
with masses as 
\bea\label{eq:LMs}
\Delta_{\mathrm{LM}\pm}=\sqrt{(-D_1\pm D_2)/(8N_F(0))},
\eea
where $D_1$ is 
\bea
D_1=4{\bf Q}^{\phi\phi}_{12}+3{\bf Q}^{\phi\phi}_{13}+3{\bf Q}^{\phi\phi}_{23},
\eea
and $D_2$ is
\bea
&D_2=\big[16({\bf Q}^{\phi\phi}_{12})^2-8{\bf Q}^{\phi\phi}_{12}{\bf Q}^{\phi\phi}_{13}+9({\bf Q}^{\phi\phi}_{13})^2\nonumber \\
&-8{\bf Q}^{\phi\phi}_{12}{\bf Q}^{\phi\phi}_{23}-14{\bf Q}^{\phi\phi}_{13}{\bf Q}^{\phi\phi}_{23}+9({\bf Q}^{\phi\phi}_{23})^2\big]^{1/2},
\eea
respectively. The two LMs are associated with the fluctuation of relative phase which is illustrated by green arrows in Fig. \ref{fig:Fig4} (a). Remarkably, LM $\Omega_{\mathrm{LM}-}$ from Eq. \ref{eq:LM-} becomes massless at the TRS breaking transition point as depicted in Fig. \ref{fig:Fig4} (a) and (b).

We further discuss fluctuations of amplitude, namely Higgs modes. It is well known that the mass of Higgs mode in single band superconductors is the twice of superconducting gap $2\Delta$, which is also the boundary of the two-quasipaticles continuum. The initial indications of Higgs mode emergence were discerned in Raman spectroscopy studies of a superconducting charge-density wave compound NbSe$_2$\cite{klein1980,varma1981gauge,varma1982}. However, within the domain of multiband superconductors, the coupling of amplitude degrees of freedom introduces intricate nuances, rendering analytical examinations challenging, both within and beyond the TRS breaking phase. Nonetheless, we navigate these complexities and numerically ascertain collective modes, encompassing both Higgs modes and Leggett modes, by solving the equation $\mathrm{Det}\hat{\bf O}=0$.

As shown in Fig. \ref{fig:Fig4} (c, d, e), the bare LM $\Omega_{\mathrm{LM}\mp}$, which is represented by the solid brown (dashed red) line and is resolved from $\mathrm{Det}{\bf O}^{\phi\phi}=0$, closely aligns with with LM $\Omega_{\mathrm{LM}\mp}$ obtained from $\mathrm{Det}\hat{\bf O}=0$. Specially, bare LM $\Omega_{\mathrm{LM}-}$ perfectly match with LM $\Omega_{\mathrm{LM}-}$ defined by the bright part of colorbar $y=-\ln|\mathrm{Det}\hat{\bf O}|$ even within TRS breaking state where $\delta\phi$ and $\delta h$ couple strongly with each other, due to the separate energy scale of Higgs mode with a heavy mass and LM $\Omega_{\mathrm{LM}-}$ with a much lighter mass.

Detailed calculations of collective modes depicted in Fig. \ref{fig:Fig4}(c, d, e) unveil that LM $\Omega_{\mathrm{LM}-}$ effectively lies well below the energy continuum of two-quasiparticle states $2\Delta_{1,2,3}$, both near or within the TRS breaking state. Its energy of this mode reaches a minimum near the transition point between different pairing states. These features underscores the potential of LM $\Omega_{\mathrm{LM}-}$ to serve as a unique character for the TRS breaking pairing\cite{xiaohu2012massless_prl}.

\section{Discussions and Conclusions}\label{sec:conclusion}

We explore the symmetry of superconducting pairing emerging from the intrinsic characteristics of the multi-Fermi surface sheets by varying pairing interactions, stemming from external pressure and doping. Including effects of the electron-phonon interaction and spin/charge fluctuations, we assume attractive intra-pocket interactions $V_{\alpha,\beta,\gamma}$. With the variation of other inter-pocket interactions $V_{\alpha\beta}$ or $V_{\beta\gamma}$, our calculations reveals a transition from $s{++}/s\pm$-wave pairing to a TRS breaking $s+i s$ state.  
Moreover, as shown in Fig. \ref{fig:Fig3}(c), we demonstrate that the variation of DOS on the $\beta$ pocket has the capacity to drive the superconducting state from a trivial $s++$ to a TRS breaking state. This TRS breaking superconducting state offers a plausible interpretation for the $\mu$SR experiments in pressurized CsV$_3$Sb$_5$\cite{Khasanov2022time,gupta2022two}. The gap amplitudes on different pockets in our calculations are uniform, which is consistent with ARPES results on doped CsV$_3$Sb$_5$ without charge order. However, gap anisotropy can occur on the hexagonal $\beta$ pocket and this leads to an anisotropic $s+is$-wave pairing. This gap anistropy goes beyond the adopted model in our work.

Furthermore, We also delve into the analysis of collective modes tied to the fluctuations in both the amplitudes and phases of the superconducting gap. Our computations reveal a remarkable outcome: a particular branch of Leggett modes becomes massless around the transition between normal $s$-wave pairing and the TRS breaking pairing. Even when phase and amplitude modes are strongly coupled in the TRS breaking state, this massless Leggett mode remains clearly separated from the two-quasiparticle continuum. Hence the LMs offer a particularly natural and sensitive probe of TRS breaking superconductivity in kagome metals, allowing one to distinguish intrinsic TRS breaking superconducting dynamics from competing TRS breaking charge orders. Despite the LM's decoupling from the electromagnetic gauge field, it can be probed through Raman spectroscopy, indirectly coupling with the electric field via its interaction with the charge density\cite{abrikosov1961zh,klein1984theory,Devereaux1995,PRL2007_observationMgB2,PRB2010_LegettMgB2,scalapino2009probing,xiaohu2012massless_prl}, and nonlinear coupling with terahertz pulse\cite{matsunaga2014light,Aoki2017theory}.

In conclusions, we uncover the intricate interplay between Fermi surface topology and electronic interactions within kagome superconductors, leading to a stabilization of a TRS breaking pairing state. A nearly massless Leggett mode emerges at the phase boundary, offering a distinctive signature.
Our work not only advances the understanding of unconventional pairing in multi-band systems but also
elucidates the collective behavior of the superconducting state, offering valuable insights into unconventional superconductivity.

\section{Acknowledgments}
We thank Jiaxin Yin for very useful discussions. We alo acknowledge the supports by the Ministry of Science and Technology  (Grant No. 2022YFA1403901), National Natural Science Foundation of China (No. 11920101005, No. 11888101,No. 12494594) and the New Cornerstone Investigator Program. F.C.Z is also partially supported by the Priority Program of Chinese Academy of Sciences (Grant No. XDB28000000). X.L.H is also supported by National Natural Science Foundation of China (Grant No. 12404162). X.W. is supported by the National Key R\&D Program of China (Grant No. 2023YFA1407300) and the National Natural Science Foundation of China (Grants No. 12574151, 12447103 and 12447101).

\bibliographystyle{apsrev4-2}

%

\end{document}